# Why scientific publications should be anonymous

Paul H. P. Hanel

Cardiff University

Numerous studies have revealed biases within the scientific communication system and across all scientific fields. For example, already prominent researchers receive disproportional credit compared to their (almost) equally qualified colleagues – because of their prominence. However, none of those studies has offered a solution as to how to decrease the incidence of these biases. In this paper I argue that by publishing anonymously, we can decrease the incidence of inaccurate heuristics in the current scientific communication system. Specific suggestions are made as to how to implement the changes.

Numerous studies have shown that humans are prone to use heuristics that make them inaccurate in their decision making (Cialdini, 2007). It can be argued that scientists are susceptible to biases based on heuristics, too. Each year thousands of articles from every scientific field are published. Given that the average amount of time a researcher spends reading is 11 hours a week (Niu & Hemminger, 2012), it is obvious that researchers need to choose what they will read. Those decisions are often based on heuristics, which, in turn, can lead to systematic biases like those addressed by Merton (1968), in his paper "The Matthew Effect in science". The Matthew effect, based on numerous interviews with Nobel laureates, can be understood as that famous researchers receive disproportional more credits than their less known, but (almost) equally qualified colleagues.

Numerous studies since the late 1960s have supported and extended the Matthew effect in science:

– Famous researchers are cited more often, even after controlling for the quality of the article (Bornmann, Schier, Marx, & Daniel, 2012; Tol, 2009).

– Researchers working at a university with a higher reputation are more likely to gain recognition (Crane, 1965) and become more often cited (Helmreich, Spence, Beane, Lucker, & Matthews, 1980).

– The so-called Matthew Effect for Countries, stating that editors and reviewers are more likely to accept – and researchers to quote – papers from specific countries like the USA, was often supported (Bonitz, Bruckner, & Scharnhorst, 1997; Link, 1998; Møller, 1990).

– Journals with high impact factors (JIF) are more likely to be cited, even after controlling for quality (Judge, Cable, Colbert, & Rynes, 2007; Larivière & Gingras, 2010; Lawrence, 2003; Opthof, 1997). Furthermore, experts tend to overrate articles published in high JIF-journals (Eyre-Walker & Stoletzki, 2013).

All of the aforementioned empirical papers are descriptive in the sense that they do not offer a strategy of how to decrease those biases. The present paper adds an important detail which has not been mentioned so far in the discussion about changes in the scientific communication (e.g., Miguel et al., 2014; Nosek & Bar-Anan, 2012): *Every publication should be published without any reference to the author(s), their position, the institution(s), the address of the researcher(s), and the journal*.

Anonymous communication would decrease the incidence of these biases and has additional advantages:

– Anonymously published arguments are more likely to become evaluated objectively (Neuroskeptic, 2013). The work of a Nobel Prize winner will only be evaluated more positive if the quality is higher, and not because she is a Nobel laureate.

– Nepotism and sexism in peer-review (Bornmann, Mutz, & Daniel, 2007; Wold & Wennerås, 1997) would be less frequent. This assumption is supported by the finding that increasing the anonymity of the involved parties decreases sexism (Budden et al., 2008).

– The so called Matilda effect, stating that the work of women in science is attributed to their male colleagues (Rossiter, 1993), would be of less importance, because it won't be possible anymore to distinguish between publications made by female and male researchers.

– Although the editors of most journals do not give the name(s) of the author(s) of a manuscript to their reviewers, it is often possible to draw inferences about the author(s) based on the publication list or the title of

Acknowledgements. I wish to thank Paul Haggar, Johannes M. Hanel, Sarah Demmrich, and Jennifer Haase for comments.

the manuscript: The double-blind review system is not yet fully established and is more like the single-blind peer-review system with all its disadvantages (Budden et al., 2008). Furthermore, editors, whose decisions are perceived as more important than those of the reviewers (Lawrence, 2003), are not blind to this information at all. If all scientific publications are submitted and research is quoted without names then the review-system will be more double-blind.

– Flattery, in the sense of citing the editors or potential reviewers in order to increase the likelihood that one's manuscript gets accepted (Seglen, 1997; Teixeira et al., 2013) will less frequent because it will be difficult for the author(s) to identify the work of editors and potential reviewers. Moreover, the work of the editors and potential reviewers will be less salient to themselves because their names are no longer mentioned.

– The fact that some authors do not cite a relevant paper because they do not want the authors of this paper as reviewers, will likely decrease as well.

– The JIF would no longer be of importance when choosing an article to read. It can be speculated that therefore the variance of the JIF of different journals would decrease and some of the negative side effects of the JIF (Lawrence, 2003) would vanish. At the moment there is likely a substantial amount of literature containing data of good quality which is missed because of the focus on journals with a high JIF (Barto & Rillig, 2012; Willmott, 2011). Ironically, the JIF is at least in ecology not even correlated with quality of the data (Barto & Rillig, 2012), and in general of dubious merit (Brembs, Button, & Munafò, 2013; Willmott, 2011).

– As names are not included, alphabetical biases (such as towards authors whose names begin with "A", Tregenza, 1997) will not occur.

One obstacle in implementing the suggested anonymous publishing system is that particularly senior researchers may feel that the credit they deserve is lost if they are not printing their names on their own work anymore. However, some suggestions are made in the next section in order to connect the name of a researcher, at least partly, with their work.

**Specific suggestions**

All of the following suggestions are meant, despite their explicit character, as a basis for discussion; to demonstrate that the aforementioned biases can likely be implemented by relatively small changes within the publishing system.
The proposed amendments could be established together with other needed changes of the scientific communication system, but they can also be implemented independently. The first step for implementing the anonymous publishing system would be to constitute a small agency to which all scientific manuscripts have to be sent before publishing. The two main tasks of the agency are (1) to replace the name of the author(s) and institutions on the manuscript with a single Digital Object Identifier (doi) and (2) forward it to the address specified by the researchers. This agency would use software calculating the citation rate: it should still be possible for each researcher to see how often her articles have been cited and to forward this list as well as selected papers to chosen institutions by email, for examples for the purpose of applications. This could be achieved automatically. The described procedure would still allow to trace the data back to the source, i.e. the fear that publishing in the suggested way would lead to more scientific fraud (Neuroskeptic, 2013) is unfounded.

*Citing.* It is suggested that, if using parenthetical referencing, we are not referring to the author(s) any more but to a few title-keywords of the article and do not refer to the journals name. Those title-keywords can replace the authors' names completely. They can be chosen by the author(s). Taking the APA citation style as an example, citing Nosek and Bar-Anan (2012) could look like this: "...how to change scientific communication (e.g. Opening Scientific Communication, 2012)". The full reference should then be placed in the reference section together with other bibliographies starting with the letter O. The latter can be applied to the Vancouver citation system, too.

*Journals.* In order to reduce the influence of the JIF on reading behavior and citation rate, it is necessary to suppress the name of a journal and to change the appearance of the journals and reduce the number of citation styles dramatically. In this way it will be much more difficult to identify a journal based on the appearance and citation style.

*Homepages.* Researchers should stop listing only their own publications on their official homepage(s). Rather, it is suggested that they list key readings in their subfield(s). As all of the references will be quoted without the author(s) name(s) it does not matter if one researcher lists 70% own publications and a second one only 25%. This ratio may vary as a function of publications of the researcher.

*Number of citations.* The total number of citations should be suppressed by web search engines and databases. However, since it is often helpful to know which articles and book(chapter)s are quoting a specific paper, it should still be possible to see the citing publications. Only a few citing articles at once should be displayed and in a random order so that it is no longer possible to estimate if a paper has been quoted 20 or 2000 times.

*Finding collaborators.* It should be possible to write private messages to the authors of the paper, which will be forwarded automatically by the repository to the contact address given by the author(s) of the paper, so that contact can be established.

*Contact with the public and (N)GOs.* The same can be done if, for example, a journalist wants to conduct an interview with a researcher. The journalist contacts the research team for a specific study through an anonymous email address given by the repository. If the researcher agrees to the interview then the topics discussed can, and need not be, limited.

*Conferences.* It would be pointless to suggest that anonymity of researchers should be maintained at conferences where researchers are presenting their work to colleagues of their own field. However, the speaker should choose a sufficiently different title from her publication.

*Book(chapter)s.* If a book(chapter) is mainly written for a scientific audience then the names of the authors, editors, publisher and the address should be removed. If a book is written for a public audience and is not relevant for scientists then there is no reason why it should be published anonymously.

*Repositories.* In line with the other changes, repositories, and social networking sites for researchers should no longer release features which emphasizes the estimated quality of the work of researchers in any way.

***Conclusion.*** The suggestions presented here can be implemented separately within the scientific communication system. Based on the doubtful worth of the JIF (Brembs et al., 2013; Larivière & Gingras, 2010; Lawrence, 2003) it seems important to suppress first the name of the journal because the bias related to the JIF is one of the strongest.